\documentclass[twocolumn]{cinc}
\usepackage{graphicx}
\usepackage{xcolor} 

\usepackage[unicode=true,bookmarks=true,bookmarksnumbered=true,bookmarksopen=true, breaklinks=true,pdfborder={0 1 1},colorlinks=true]{hyperref}
\begin{document}

\bibliographystyle{cinc}

% Keep the title short enough to fit on a single line if possible.
% Don't end it with a full stop (period).  Don't use ALL CAPS.
\title{%Investigation of Poor Neuro-prognostication in Comatose Cardiac Arrest Patients: A Representative Study using Attention Mechanism
A Multi-channel EEG Data Analysis for Poor Neuro-prognostication in Comatose Patients with Self and Cross-channel Attention Mechanism
}
% Both authors and affiliations go in the \author{ ... } block.
% List initials and surnames of authors, no full stops (periods),
%  titles, or degrees.
% Don't use ALL CAPS, and don't use ``and'' before the name of the
%  last author.
% Leave an empty line between authors and affiliations.
% List affiliations, city, [state or province,] country only
%  (no street addresses or postcodes).
% If there are multiple affiliations, use superscript numerals to associate
%  each author with his or her affiliations, as in the example below.
\author {Hemin Ali Qadir$^{1}$, Naimahmed Nesaragi$^{1}$\thanks{Corresponding Author: Naimahmed Nesaragi,\\Email Address: naimahmed.nesaragi@gmail.com}, Per Steiner Halvorsen$^{1, 2}$, Ilangko Balasingham$^{1, 3}$\\
\ \\ % leave an empty line between authors and affiliation
 $^1$ The Intervention Centre, Oslo University Hospital, Oslo, Norway \\
 $^2$ Institute of Clinical Medicine, Faculty of Medicine, University of Oslo, Oslo, Norway \\
 $^3$ Dep. of Electronic Systems, Norwegian University of Science and Technology, Trondheim, Norway %\\
%$^2$  Second Institution, Other City, Country  
}

\maketitle
\renewcommand\footnoterule{%
  \kern-3pt
  \hrule width 1.0\columnwidth
  \kern2.6pt
}

\begin{abstract}
This work investigates the predictive potential of bipolar electroencephalogram (EEG) recordings towards efficient prediction of poor neurological outcomes. A retrospective design using a hybrid deep learning approach is utilized to optimize an objective function aiming for high specificity, i.e., true positive rate (TPR) with reduced false positives ($\leq$ 0.05). A multi-channel EEG array of 18 bipolar channel pairs from a randomly selected  5-minute segment in an hour is kept. In order to determine the outcome prediction, a combination of a feature encoder with 1-D convolutional layers, learnable position encoding, a context network with attention mechanisms, and finally, a regressor and classifier blocks are used. The feature encoder extricates local temporal and spatial features, while the following position encoding and attention mechanisms attempt to capture global temporal dependencies. Results: The proposed framework by our team, \textbf{OUS\_IVS}, when validated on the challenge hidden validation data, exhibited a score of \textbf{0.57}. The GitHub link goes here. 
\\

{
}

\end{abstract}
\section{Introduction}
Assessment of integral neurological outcomes in comatose patients after cardiac arrest is an ongoing scientific challenge to the clinical world. The current prognostication methods on comatose patients with the return of heart function, i.e. return of spontaneous circulation, primarily rely on subjective visual expert scoring of physiological signals. However, this approach is susceptible to inherent subjectivity, leaving a significant number of patients categorized as an ambiguous "grey zone" with uncertain prognosis. The general rule of thumb in practice is to provide continuous care for a favorable outcome (perceived good prognosis) while withdrawing the life-sustaining therapy for a perceived poor prognosis, thereby causing death. However, uncertainty in subjective prognostication has resulted in increased false-positive rates (cases where patients have made a good recovery despite a perceived poor prognosis). This raises concerns about potential self-fulfilling prophecies and premature withdrawal of life-sustaining therapy (WLST). 

Multi-channel EEG streams aid in reducing the subjectivity of prognostic evaluation, and several prognostic indicators are already been identified based on the considered outcome (good/poor prognosis) following cardiac arrest\cite{sandroni2022prediction,rossetti2017electroencephalography,bongiovanni2020standardized,ruijter2019early}. Burst suppression, nonreactive patterns or isoelectric EEG indicates poor prognosis, but the interpretable quantification of EEG streams is a laborious task that demands advanced clinical and neurophysiological expertise, inhibiting the accessibility of EEG-informed prognostication. Automating EEG interpretation has the potential to improve accessibility and diagnostic accuracy.

\begin{figure*}[!t]
\centering
\includegraphics[scale=0.55]{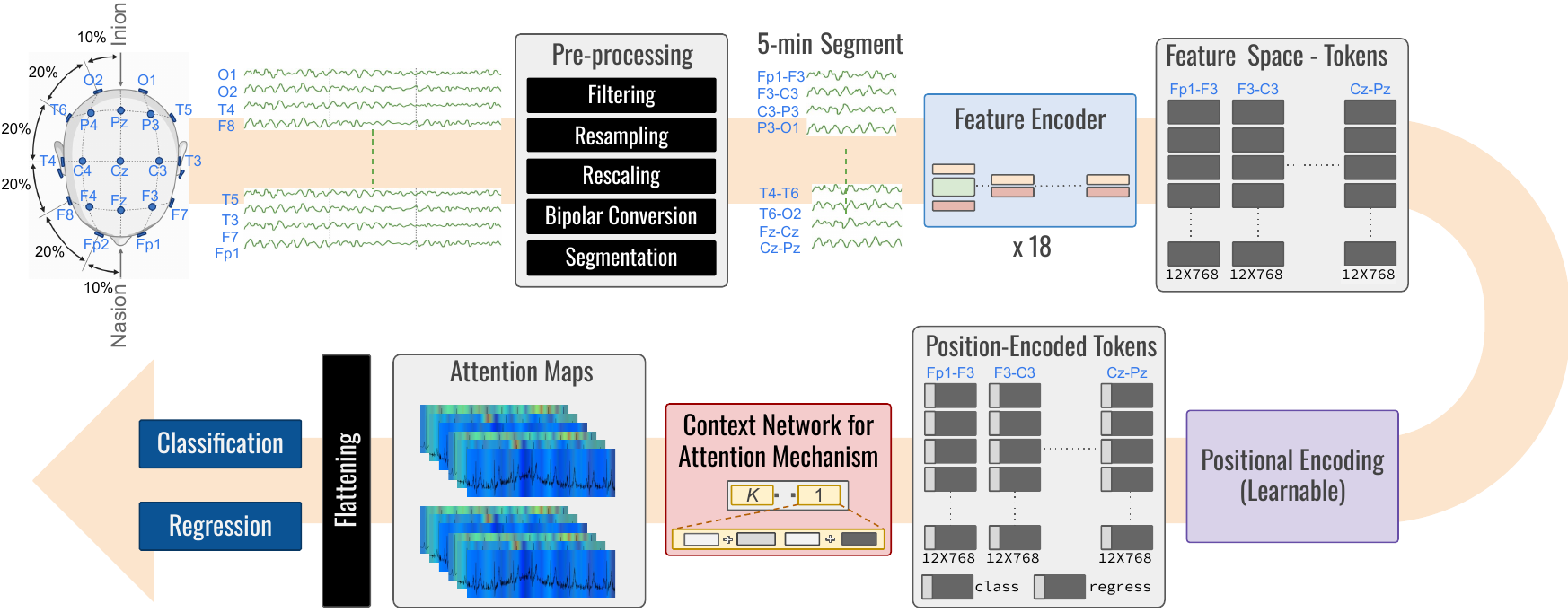}
\caption{The proposed framework: A raw multi-channel EEG data is fed as initial input to the \textit{Pre-processing} pipeline where the data is filtered, resampled, rescaled, segmented, and converted to bipolar signals. A pre-processed 5-minute segment is fed to \textit{Feature Encoder} which converts each bipolar channel input to a lower-dimensional feature representation called \textit{Tokens} preserving local patterns and spatial relationships. The information about the position of each token in the sequential feature space is encoded by the \textit{Positional Encoding} block. The \textit{Context Network} enables the framework to focus on relevant information while considering the long-range relationships between inter- and intra- channels. The resulting attention maps are flattened and fed to \textit{Classification and Regression} blocks for binary classification and CPC prediction.}
\label{FIGURE1}
\vspace{-1.0em}
\end{figure*}

In recent years, deep learning (DL)-based attention mechanisms have presented an intriguing avenue for further exploration of the multi-channel integration of the brain \cite{shi2022multimodal}. The attention mechanisms enable DL models to focus on relevant information while considering the long-range relationships among different parts of inter and intra-channel signals. The attention mechanism has shown remarkable success in natural language processing to capture long-range dependencies compared to convolutional neural networks (CNNs). In this study, we hypothesize that attention mechanisms can enhance the interpretability and predictive power of multichannel EEG data to classify comatose patients with good or poor neurological outcome. Specifically, we propose that attention mechanisms can effectively discern both self-attention patterns within individual EEG channels and cross-attention patterns among multiple EEG channels, thus providing critical insights into distinct brain activities underlying the comatose state, leading to improved classification accuracy. 
%In recent state-of-the-art artificial intelligence (AI)  methods focusing on automated analysis, attention-mechanism-based deep learning methods present an intriguing avenue for further exploration of the multi-channel integration mechanisms of the brain \cite{shi2022multimodal}. The attention mechanism enables the model to focus on relevant information while considering the relationships between inter and intra-channel modalities. Originally introduced in NLP tasks, the attention mechanism has shown remarkable success in capturing long-range dependencies compared to CNNs. Applying the attention mechanism to multi-channel data fusion could enhance the ability to integrate and analyze diverse information sources, improving diagnostic accuracy and prognostication reliability. This study proposes a representative idea aimed at uncovering the potential of \textit{attention mechanism} to catalyze a paradigm shift in assessing neurological outcomes in comatose patients.
\section{Materials and Methods}
\vspace{-1.0em}
\subsection{Pre-processing}\label{preprocess}
\vspace{-0.7em}
The data pre-processing pipeline includes the following sequence of operations: filtering, re-sampling, re-scaling, bipolar conversion, and finally, segmentation of  EEG recordings. At the onset, the entire EEG data is filtered using a Butterworth band pass filter with cut-off frequencies of 0.5 Hz and 35 Hz to remove baseline wander and high-frequency noises from the EEG signals. Next, the filtered signals are examined in terms of sampling frequency, and all the EEG recordings are re-sampled to 100 Hz to maintain uniformity with respect to sampling frequencies. The re-sampled signals are then re-scaled using Min-max standardization. Next, re-scaled signals are converted to bipolar representations and finally segmented into 5-minute segments. The bipolar conversion, subtracting EEG signals from adjacent scalp electrodes, is crucial in EEG signal processing. It reduces noise and artifacts, enhances spatial resolution by focusing on localized brain activity, minimizes volume conduction effects, and aids comparisons to baseline states. Valuable in clinical and neuroscience research, it provides a cleaner and more accurate brain activity representation \cite{kutluay2019montages}. This data pre-processing pipeline yields a set of 5-minute segments of 18 bipolar channels from every hour of EEG recordings. 
\subsection{Framework}
\vspace{-0.75em}
Figure \ref{FIGURE1} depicts the proposed framework's design, with further sections explaining each block in detail.

%A pre-processed 5-minute segment of 18 bipolar EEG channels is fed as initial input to the framework. Each bipolar channel input is further subjected to a tokenization by a CNN embedding block called the \textit{Feature Encoder}. This embedding block transforms the 18 bipolar EEG channels into a lower-dimensional feature representation called \textit{Tokens}. The information about the position or the order of each token in the sequential feature space is encoded by the \textit{Positional Encoding} block. The attention mechanism on the resultant position-encoded tokens is provided by the subsequent \textit{Context Network}. The context network enables the framework to focus on relevant information while considering the relationships between inter and intra-channels.

\subsubsection{Feature Encoder} 
\begin{figure}[!h]
    \vspace{-0.5em}
    \centering
    \includegraphics[scale=0.52]{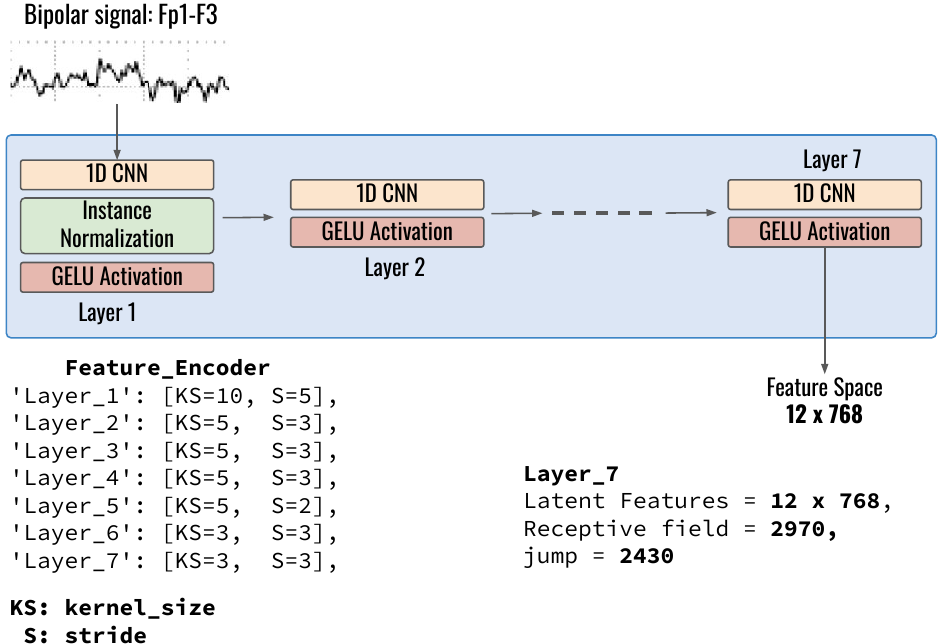}
    \caption{Feature Encoder.}
    \label{fig:Featureencoder}
    \vspace{-2em}
\end{figure}

\vspace{-0.7em}
In this study, we employ a systematic approach to handle the pre-processed 5-minute segments of 18 bipolar EEG channels. Each bipolar channel segment is passed through its own dedicated feature encoder block, resulting in a total of 18 such blocks. Each feature encoder block consists of a stack of seven 1D CNN layers, as illustrated in Figure \ref{fig:Featureencoder}. The initial layer includes an instance normalization between the 1D convolution and the Gaussian Error Linear Unit (GELU) activation function. The other six layers consist of a 1D convolution followed by a GELU activation function. The total context of the encoder receptive field at $7^{th}$ layer is 2970 samples with a jump of 2430, corresponding to $\sim$30 seconds at the 100 Hz sample rate. Hence, the feature encoder blocks converted $\sim$30-second fragment to a token. As a result, 12 tokens are generated from each bipolar channel from a 5-minute segment. The output of the feature encoder yields a feature vector with dimensions $12\times768$ per bipolar channel and a total feature space dimension of $216\times768$ from the multi-channel EEG array of 18 bipolar channels.  
This stack of seven 1D CNN layers excels at capturing local patterns and dependencies within the EEG channels with respect to time, effectively reducing the data dimensionality while preserving relevant information. This feature extraction process transforms raw EEG signals into compact representations in the form of tokens that the context network of the attention mechanism block can further process. Tokenization allows the EEG data to be organized into discrete fragments, enabling the attention mechanism to model long-range dependencies and capture complex relationships within inter- and intra- EEG channels.
%Applying 1D CNNs on EEG data serves as a crucial step to extract meaningful features from the complex temporal patterns present in the brain signals. 

\subsubsection{Positional Encoding}
\vspace{-1em}
Positional encoding adds encoding vectors to the token embeddings, i.e., it encodes information about the position of each token. The significance of positional encoding lies in enabling an attention mechanism (see subsection \ref{Attention Mechanism}) to handle sequences of variable length while preserving their order. In the proposed framework, the positional encoding block passes the resultant feature space through learnable positional vectors, which are learned and then appended as addresses to encode the positions of the tokens relative to one another during the training phase. We use a learnable positional encoding scheme over other approaches because it offers flexibility in modeling complex position-dependent relationships, where the patterns in data might not conform to simple sinusoidal functions, and hence, there is a need to capture more nuanced position-based information. Further, if the input sequences have varying lengths, learnable positional encodings can adapt to these variations, ensuring that positional information remains meaningful regardless of sequence length. It is worth mentioning that we prepend learnable $[class]$ and $[regress]$ tokens to the resulting feature space (see Figure \ref{FIGURE1}), leading to the dimension of $218\times768$. The state value of these two tokens serves as the class and regression representations in the feature space.

\subsubsection{Context Network}\label{Attention Mechanism}
\begin{figure}[!h]
    \centering
    \includegraphics[scale=0.5]{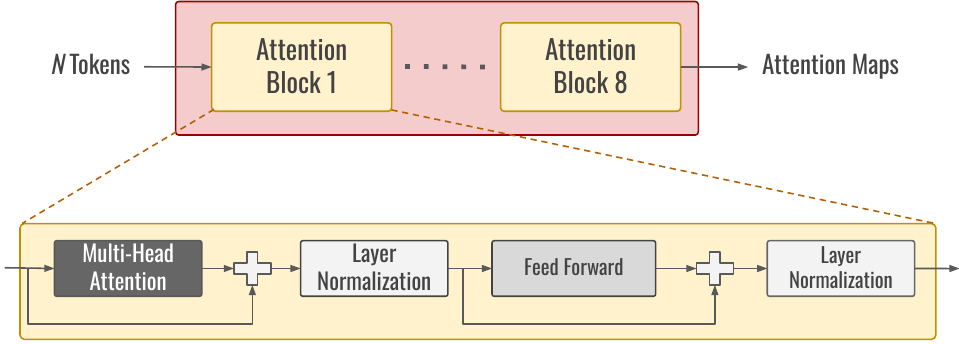}
    \caption{Context Network for Attention Mechanism}
    \label{fig:attention}
    \vspace{-0.5em}
\end{figure}

\vspace{-0.5em}
As shown in Figure \ref{fig:attention}, the context network of the attention mechanism comprises $K$ attention blocks in succession. Each attention block applies multi-head attention with $M$ heads and a subsequent feed-forward network, both followed by layer normalization. The feature encoder uses convolutional filters with limited receptive fields, with 2970 samples designated for the last layer. This yields the feature encoder suitable for capturing local patterns and short-term dependencies. However, EEG multi-channel data analysis can benefit from global patterns and long-range dependencies across various time steps. We address this by proposing a hybrid architecture combining 1D CNNs and attention mechanisms. This hybrid approach aids in modeling and extracting features from EEG data with both short-term and long-term dependencies. Using our context network, we aim to capture hierarchical temporal dependencies by combining short-term patterns captured by the feature encoder to form longer-term patterns. Such a design seeks to model the interaction between patterns at different time scales. The resulting feature space contains tokens created from 18 bipolar EEG channels. This token fusion approach facilitates the context network to model long-range dependencies and capture complex relationships within inter- and intra-EEG channels simultaneously without adding complexity to the model design. 

%The context network enables not only a channel-wise attention mechanism but also accomplishes a cross-channel attention mechanism on the resultant feature space of 216 x 768 tokens from multi-channel EEG data.                          
%The attention mechanism enables the framework to focus on relevant parts of the input during the processing.

%The context network leverages these 5-minute segments to discover the complex short- and long- range dependencies present within approximately 30-second fragments within each 5-minute multi-channel segment, as mentioned in Subsection \ref{preprocess}. 

\subsubsection{Classification and Regression}
\vspace{-0.5em}
The context network is followed by two blocks: the classification and regression blocks. Both these blocks consist of a single layer of fully connected neurons. Once the attention maps are obtained, they will be flattened into a vector that these two blocks can handle. The classification block attempts to forecast the neurological outcome, which is a binary decision: “0” for a good outcome or “1” for a poor outcome. In contrast, the regression block aims to predict the Cerebral Performance Category (CPC) value, which is an ordinal scale ranging from 1 to 5 \cite{arawwawala2007clinical}.

\subsection{Training Details}
\vspace{-0.5em}
Attention mechanisms typically require a substantial amount of training data to capture inherent correlations effectively. This is accomplished by adapting a segmentation approach for the selected hour of EEG recording into 5-minute segments. This ensures the generation of a sufficient number of training samples to support efficient end-to-end training of our context network. To further enhance training robustness, each iteration is started by randomly selecting an hour of EEG recording of a patient, followed by a segmentation strategy, and then selecting a random 5-minute segment from the corresponding random hour. Using this random selection and 5-minute segmentation techniques, we could generate $\sim$0.5 million training samples from the under-studied dataset from EEG recordings of 607 patients (See Section \ref{Results}).

We performed a patient-stratified data split on a given training dataset, resulting in two in-house subsets: an 80\% training set and a 20\% validation set, ensuring that the patient's data remained intact within each subset. The in-house validation set served as a means for frequently assessing the trained model's performance, and the best model, determined by the highest accuracy, was retained. Adam optimizer is employed to iteratively update the model's weights, with a batch size of 10 and a learning rate of 0.0001. The training process spanned across 40,000 epochs. For the classification task, we adapted the cross-entropy loss function, tailored to the 'Good' and 'Poor' neurological outcome labels: \vspace{-0.75em}\begin{equation}
L_{(\hat{y}, y)} = -\frac{1}{N}\sum_{i=1}^{N} \left[ y_i \log(\hat{y}_i) + (1-y_i) \log(1-\hat{y}_i) \right]
\end{equation} where \(N\) is the batch size, \(y_i\) is the true label (\(y_i = 1\) for 'Good' and \(y_i = 0\) for 'Poor'), and \(\hat{y}_i\) is the predicted outcome for the \(i\)-th patient. For the regression task, we utilized the Mean Squared Error (MSE) loss function to predict the CPC values: \vspace{-0.75em}\begin{equation}
L_{(MSE)} = \frac{1}{N}\sum_{i=1}^{N} (x_i - \hat{x}_i)^2
\end{equation} where \(x_i\) stands for the true CPC value, and \(\hat{x}_i\) is the predicted CPC value for the \(i\)-th patient. The total loss for the model is the sum of these two losses, encompassing both classification and regression objectives:\vspace{-0.75em}\begin{equation}
L_{(total)} = L_{(\hat{y}, y)} + L_{(MSE)} 
\end{equation} %This combined loss guides the training process, enabling the model to simultaneously handle classification and regression tasks.

\section{Results and Discussion}
\label{Results}
The data under study is provided by the PhysioNet/CinC Challenge 2023 assembled by the International Cardiac Arrest Research Consortium (I-CARE). I-CARE data includes seven hospitals in the United States and Europe. The training and validation sets have patient data, including EEG records, from 5 hospitals: A, B, D, E, and F, while the test set has records from an extra hospital, C. The exclusion of hospital C from the training and validation sets helps to assess the generalizability of the methods. The training set contains records from 607 patients ($\sim$60\%), the validation set has $\sim$10\%, and the test set has $\sim$30\% of 1020 total patient records.

Further, the performance of our framework was evaluated officially on the challenge validation dataset and obtained a score of 0.57. Table \ref{tab:font1} details the performance of various entries submitted. Our team, \textbf{OUS\_IVS}, ranks \textbf{64} in the official leaderboard. Our initial exploration focused on the last 5-minute segment of the last hour using only 2 bipolar channels with 2 attention blocks each comprising 2 heads. This resulted in our $1^{st}$ entry yielding a low score of 0.09 \begin{table}[ht!] 
\caption{\label{tab:font1} Challenge metric for various entries.}
\vspace{3 mm}
\small
\centering
%\renewcommand{\arraystretch}{1.8}
%\resizebox{1.0\columnwidth}{1cm}
{\begin{tabular}{l|c|c|c|c} \hline \hline
 Entry & \# of bipolar  & \# of Attention  & \# of & Challenge  \\ No. & channels & Blocks & Heads & Metric  \\ \hline
1. & 2     & 2  & 2 & 0.09  \\ 
2. & 2     & 2  & 2  & 0.45 \\
3. & 2     & 8  & 8 & 0.54 \\
4. & 18     & 8  & 8 & 0.57 \\
\hline
\end{tabular}}
\end{table}due to over-fitting. So, our $2^{nd}$ entry focused on the random 5-minute segment of the retained random hour, while keeping the same framework design. This improved our results to 0.45, indicating the elimination of over-fitting. Next, we tried to increase the effect of the attention mechanism from 2 to 8 attention blocks, each with 8 heads. This resulted in our $3^{rd}$ entry with further enhancement of the challenge score to 0.54. Finally, we incorporated the 18 bipolar channel pairs. This resulted in our $4^{th}$ entry with best score of 0.57.

\section{Conclusion}
This study has shown that a hybrid approach combining 1D CNNs and attention mechanisms can discern both local and global self- and cross-attention patterns within single and multiple EEG channels to enhance the interpretability and predictive power, leading to efficient binary classification of comatose patients into neurological outcomes.

%Specifically, we want a high degree of certainty, i.e. \textit{“When the prognosis is poor, continued treatment would not yield a good outcome.”} The goal is to reduce false positives in case of poor prognosis and mitigate potential self-fulfilling prophecies and premature WLST decisions based on uncertain prognostic information.

\bibliography{refs}

\end{document}